\documentclass[twocolumn]{aastex62}
\usepackage{amsmath}

\accepted{April 16, 2021}
\submitjournal{ApJS}
\usepackage{multirow}

\begin{document}

\title{The Lithium Abundances from the Large Sky Area Multi-object Fiber Spectroscopic Telescope Medium-resolution Survey. I. The Method}

\author[0000-0003-4972-0677]{Qi Gao}
\altaffiliation{}
\affil{CAS Key Laboratory of Optical Astronomy, National Astronomical Observatories, Beijing 100101, China}
\affil{School of Astronomy and Space Science, University of Chinese Academy of Sciences, Beijing 100049, People’s Republic of China}

\author[0000-0002-0349-7839]{Jian-rong Shi}
\altaffiliation{E-mail: sjr@bao.ac.cn}
\affil{CAS Key Laboratory of Optical Astronomy, National Astronomical Observatories, Beijing 100101, China}
\affil{School of Astronomy and Space Science, University of Chinese Academy of Sciences, Beijing 100049, People’s Republic of China}

\author[0000-0002-8609-3599]{Hong-liang Yan}
\altaffiliation{E-mail: hlyan@nao.cas.cn}
\affil{CAS Key Laboratory of Optical Astronomy, National Astronomical Observatories, Beijing 100101, China}
\affil{School of Astronomy and Space Science, University of Chinese Academy of Sciences, Beijing 100049, People’s Republic of China}

\author[0000-0002-6647-3957]{Chun-qian Li}
\affil{CAS Key Laboratory of Optical Astronomy, National Astronomical Observatories, Beijing 100101, China}
\affil{School of Astronomy and Space Science, University of Chinese Academy of Sciences, Beijing 100049, People’s Republic of China}

\author[0000-0002-6448-8995]{Tian-Yi Chen}
\affil{CAS Key Laboratory of Optical Astronomy, National Astronomical Observatories, Beijing 100101, China}
\affil{School of Astronomy and Space Science, University of Chinese Academy of Sciences, Beijing 100049, People’s Republic of China}

\author[0000-0002-2510-6931]{Jing-Hua Zhang}
\affil{CAS Key Laboratory of Optical Astronomy, National Astronomical Observatories, Beijing 100101, China}

\author[0000-0001-5193-1727]{Shuai Liu}
\affil{CAS Key Laboratory of Optical Astronomy, National Astronomical Observatories, Beijing 100101, China}
\affil{School of Astronomy and Space Science, University of Chinese Academy of Sciences, Beijing 100049, People’s Republic of China}

\author[0000-0003-2318-9013]{Tai-Sheng Yan}
\affil{CAS Key Laboratory of Optical Astronomy, National Astronomical Observatories, Beijing 100101, China}
\affil{School of Astronomy and Space Science, University of Chinese Academy of Sciences, Beijing 100049, People’s Republic of China}

\author{Xiao-Jin Xie}
\affil{CAS Key Laboratory of Optical Astronomy, National Astronomical Observatories, Beijing 100101, China}
\affil{Tibet University, Lhasa 850000, People’s Republic of China}

\author{Ming-Yi Ding}
\affil{CAS Key Laboratory of Optical Astronomy, National Astronomical Observatories, Beijing 100101, China}
\affil{School of Astronomy and Space Science, University of Chinese Academy of Sciences, Beijing 100049, People’s Republic of China}

\author[0000-0003-2179-3698]{Yong Zhang}
\affil{Nanjing Institute of Astronomical Optics and Technology, National Astronomical Observatories, Chinese Academy of Sciences, Nanjing 210042, People’s Republic of China}

\author[0000-0002-3701-6626]{Yong-Hui Hou}
\affil{School of Astronomy and Space Science, University of Chinese Academy of Sciences, Beijing 100049, People’s Republic of China}
\affil{Nanjing Institute of Astronomical Optics and Technology, National Astronomical Observatories, Chinese Academy of Sciences, Nanjing 210042, People’s Republic of China}

\begin{abstract}
One of the purposes of taking spectra for millions of stars through the Large Sky Area Multi-Object Fiber Spectroscopic Telescope (LAMOST) medium-resolution survey (MRS) is to obtain the elemental abundances, so that one can trace the origin and evolution for the element of interests. Lithium is one of such elements of great importance, which exhibits many puzzling behaviors. Investigating the lithium abundances to a uniquely large sample of stars is essential to understand its origin and evolution. In this paper, we present the lithium abundances obtained from the LAMOST MRS spectra calculated by the template-matching method. Our dataset consists of $294,857$ spectra that corresponds to $165,479$ stars with a resolution power of $\Delta \lambda/\lambda \sim 7,500$. We compared the lithium abundances derived from our work with those using the high-resolution spectra and found a good consistence. The errors of lithium abundances are discussed. Our results suggest that the distribution of lithium abundances show two clear peaks at $+2.6$\,dex and $+1.0$\,dex, respectively. This sample is potentially important for investigating physical mechanisms occurring inside stars that alter the surface lithium abundance.

\end{abstract}


\section{INTRODUCTION}

Lithium is one of the most interesting elements, and shows complex pattern in different types of stars, which are not well understood. 

The primordial lithium was synthesized during the first twenty-minutes of the universe. Based on the baryon density inferred from the Cosmic Microwave Background observations or from the primordial deuterium measurements, the standard big bang nucleosynthesis (SBBN) predicts that the primordial lithium abundance is $\sim$ 2.7\,dex \citep{Cyburt2003, Cyburt2008, Cyburt2016, Coc2012, Coc2014, Fields2020, Spergel2003}, which has a significant discrepancy with the observed value. \citet{Spite1982} and the following works \citep{Bonifacio1997, Asplund2006, Bonifacio2007, shi2007, Molaro2008, Lind2009, Sbordone2010} confirmed that most of the warm (\textgreater 5,700\,K) metal-poor ($-2.4$\textless [Fe/H]\textless$-$1.4) main-sequence (MS) stars display a plateau of the Li abundance (the so-called `Spite Plateau'), sharing a similar value at A(Li) $\sim$ 2.2\,dex. This discrepancy is generally denominated as the $\emph{cosmological lithium problem}$. Many studies have been devoted to solve this problem \citep{Coc2012, Olive2012, Kajino2012, Kusakabe2015, hou2017, kang2019, korn2020, Sasankan2020}, however, it is still unsolved \citep[see a recent review,][]{mathews2020}.

For the Milky Way disk, the increase in measurements provided by recent works indicated that there is a different behaviour in lithium evolution between the thin and thick disk stars. At the lower metallicity end, they present the same lithium abundances, while, the increase in lithium abundances is steeper in thin than in thick disks at higher metallicities \citep{Molaro1997, Guiglion2016, Fu2018}. The behaviour of lithium abundances in the thick disk as a function of metallicity is still in debate. 

Also, it is found that, for the disk stars, the upper envelope of lithium abundances increase above the Spite Plateau level at higher metallicities \citep{Ram2012,Delgado2015, Guiglion2016, Fu2018}, which indicates that the Galaxy has undergone a history of lithium enrichment since the Big Bang. However, the sources of enrichment are not well understood, which is referred as the \emph{Galactic lithium problem}. Several sources have been proposed: a). The spallation of ISM atoms by high-energy cosmic rays, which accounts for less than 30$\%$ of the meteoritic lithium abundance \citep{Reeves1970, Meneguzzi1971, Lemoine1998, Romano2001, Prantzos2012}; b). The neutrino process in the core-collapse supernovae, although no lithium lines have been found from the supernovae's spectra \citep{Domogatskii1978, Hartmann1999, Yoshida2005, Yoshida2004}; c). Classical novae explosions during which $^{7}$Li \citep{Izzo2015} or $^{7}$Be lines (will decay to $^{7}$Li in a short time) have been detected \citep{Tajitsu2015, Tajitsu2016, Molaro2016, Selvelli2018}; d). The evolved stars, which are at the red giant branch (RGB), the red clump (RC), or the asymptotic giant branch (AGB) phases. It is found that some evolved stars \citep[e.g.][and references therein]{Yan2018, Casey2019, Gao2019, Singh2019, Kumar2020, Yan2021, Smith1989, Smith1990, Abia1999} have lithium abundances even higher than the meteoritic value.

Above the solar-metallicities ([Fe/H] \textgreater 0), lithium abundances in solar neighborhood stars show an puzzling decrease \citep{Delgado2015, Guiglion2016, Bensby2018, Buder2018, Fu2018}. Similar situation has been found for the dwarfs and subgiants in Galactic bulge \citep{Bensby2020}. \citet{Stonkute2020} presented that, at the supersolar-metallicities, the lithium abundances descended by 0.7\,dex in the [Fe/H] range from $+$0.10 to $+$0.55\,dex. However, \citet{Randich2020} noted that their solar- and supersolar-metallicity stars all have higher lithium values with a peak at $\sim +3.4$\,dex for the two most metal-rich open clusters, thus, they suggested that the previous findings based on field stars were biased by selection effects.

The depletion of  lithium on the surface of a star makes the studies of the phenomenon on the lithium abundances more complicated. Lithium will be destroyed at relatively low temperatures in stellar interiors \citep{2019Deliyannis}, and the Li destruction can be observed at the photosphere if standard convective mixing reaches down as far as the Li-burning regions. In pre-MS and MS stars, considerable evidence suggests that the dominant mechanism depleting surface lithium over time is the rotational induced mixing \citep{1998Boesgaard, 2005Boesgaard, 2016Boesgaard, 2020Boesgaard, Cummings2017, 1994Deliyannis, 1998Deliyannis}. While, in subgiant and giant stars, the surface lithium may be consumed more severely due to the deepening of the convective zone, rotational mixing and/or thermohaline mixing \citep{2018Anthony, Anthony-Twarog2021, 1999Charbonnel, 2010Charbonnel, 1990Deliyannis, 1995Ryan}.

In order to reveal the lithium behaviors, a large and homogeneous sample with precise lithium abundances is needed. Such efforts have been attempted in recent spectroscopic surveys, such as the Gaia-ESO survey \citep[e.g.][]{Fu2018, Randich2020}, the AMBRE project \citep[e.g.][]{Guiglion2016, Prantzos2017}, Galactic Archaeology with HERMES (GALAH) survey \citep[e.g.][]{Zerjal2019, Deepak2020, Gao2020}, etc. The Large Sky Area Multi-Object Fiber Spectroscopic Telescope \citep[LAMOST,][]{Cui2012, Zhao2012} launched the medium-resolution ($\Delta \lambda/\lambda$ $\simeq 7,500$) spectroscopic survey (hereafter MRS) \citep{Liu2020} in September 2018. Its data are potentially important for lithium abundance investigation in the sense that it has already obtained massive high-quality MRS spectra that cover lithium resonance line in the red band.

The purpose of this paper is to present the derived lithium abundances for 294,857 spectra of 165,479 targets from the LAMOST MRS spectra. Our paper is organized as follows: Section \ref{sec:DATA} describes the spectra and stellar parameters we used, and Section \ref{sec:METHOD} presents the method we developed for deriving lithium abundances. Validation of the method and error estimation are provided in Section \ref{sec:Validation} and Section \ref{sec:Error}, respectively. The resulting catalogue and the distributions of the lithium abundances are shown in Section \ref{sec:RESULT}. Finally, a brief summary is given in Section \ref{sec:SUMMARY}.

\section{DATA}\label{sec:DATA}

\begin{figure*}[!t]
\includegraphics[scale=0.59]{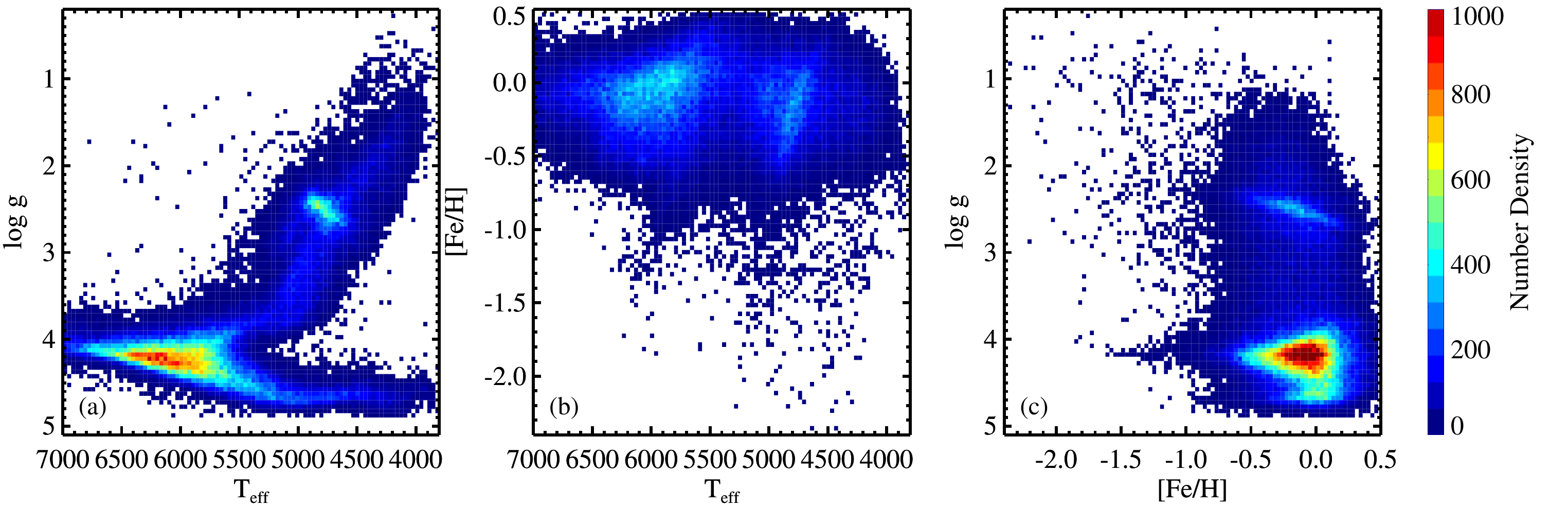}
\centering
\caption{Distributions of our sample stars in the planes of $T_{\rm eff}$-$\log{g}$ (a), $T_{\rm eff}$-[Fe/H] (b) and [Fe/H]-$\log{g}$ (c). The colors indicate the number densities of stars.} 
\label{fig:1}
\end{figure*}

\begin{figure*}[!t]
\includegraphics[scale=0.44]{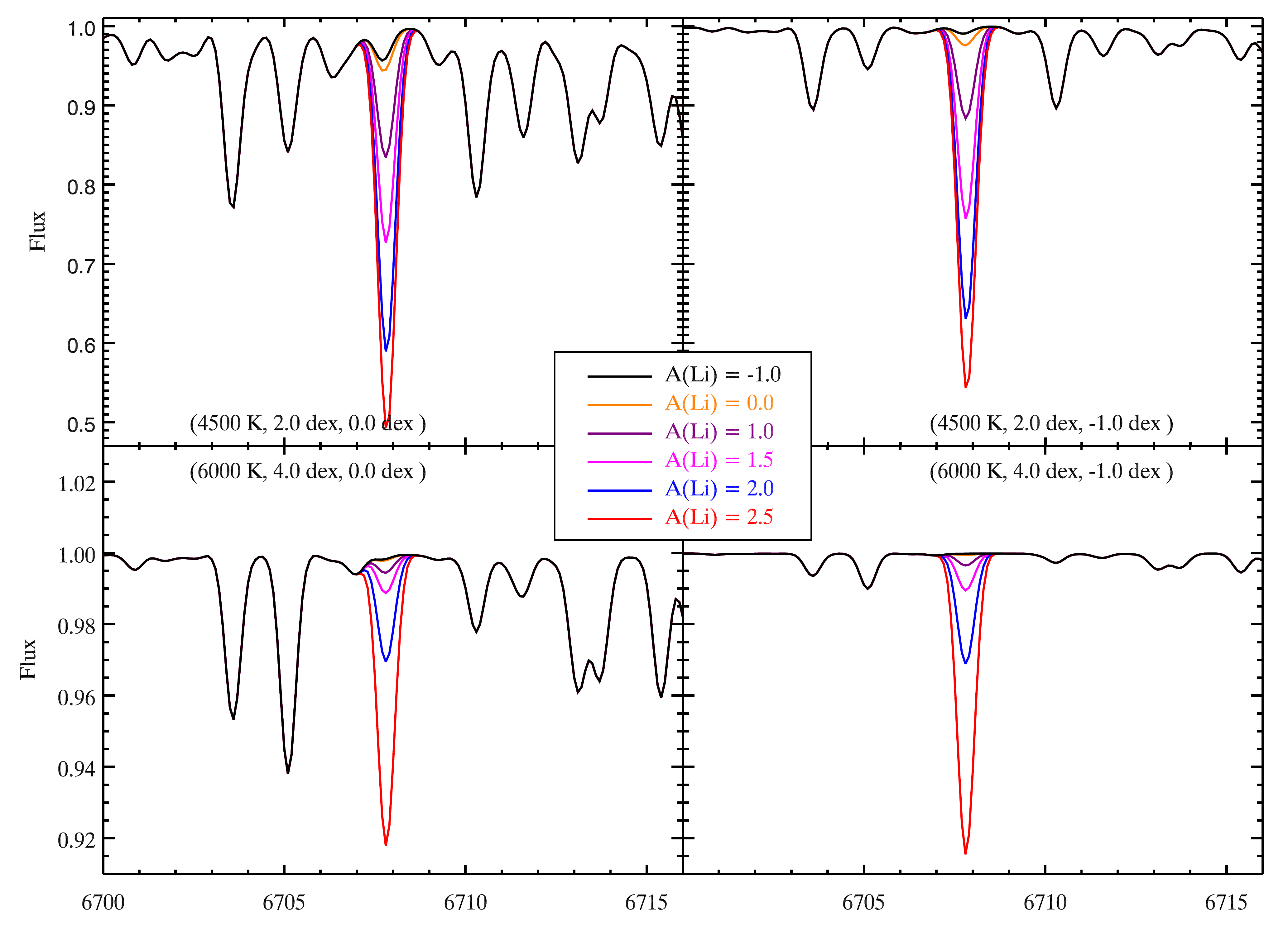}
\caption{ Theoretical \ion{Li}{1} resonance lines at 6708\,\AA\ with four sets of stellar parameters, and A(Li) varying from $-$1.0 to $+$2.5 are shown with different colors. The stellar parameters ($T_{\rm eff}$, $\log{g}$ and [Fe/H]) are indicated.
\label{fig:2}}
\end{figure*}

\subsection{Spectra} 

The LAMOST MRS spectra consist of the blue and red band, and the wavelength ranges are 4,950 $\sim$ 5,350\,\AA\ and 6,300 $\sim$ 6,800\,\AA, respectively \citep{Liu2020}. In our study, the red band spectra observed from September 2017 to June 2019 with a signal-to-noise ratio (S/N) higher than 10 have been selected\footnote{http://dr7.lamost.org/}.

\subsection{Stellar parameters} 

The effective temperatures ($T_{\rm eff}$), surface gravities ($\log{g}$), metallicities ([Fe/H]) and micro-turbulent velocities ($\xi$) of our sample stars have not been well determined from the MRS spectra at the moment, thus we adopted the stellar parameters from the following works:

--- The stellar parameters of LAMOST DR7: they were derived from the LAMOST low-resolution spectra with the LAMOST Stellar Parameter Pipeline \citep[LASP,][]{Luo2015} using a template matching method, and the template spectra are produced based on the ELODIE spectra library \citep{Prugniel2001, Prugniel2007}. This parameter dataset covers 267,962 spectra of 150,553 stars in our sample (hereafter the LASP-SP).

--- The stellar parameters of SDSS DR16: they were from the Apache Point Observatory Galactic Evolution Experiment (APOGEE)-2/Sloan Digital Sky Survey \citep[SDSS-IV survey,][] {Jonsson2020} determined by the APOGEE Stellar Parameter and Chemical Abundances Pipeline \citep[ASPCAP,][]{Garca2016}, this method matches the observed spectra to the pre-computed theoretical templates calculated with a local thermodynamic equilibrium (LTE) assumption. This parameter dataset covers 26,895 spectra of 14,926 stars in our sample (hereafter the APOGEE-SP).

For the stars have both LASP and APOGEE parameter, the latter is adopted, as APOGEE is a high resolution survey. 

Finally, our sample consists of 294,857 spectra of 165,479 targets, spanning a range of stellar atmospheric parameters: 3,500\,K$\textless T_{\rm eff} \textless$7,000\,K, 0.1\,dex$\textless \log{g} \textless$5.0\,dex, $-$2.5\,dex$\textless$ [Fe/H]$\textless+0.5$\,dex. Figure \ref{fig:1} presents the stellar parameter spaces of our sample in the planes of $T_{\rm eff}$-$\log{g}$ (a), $T_{\rm eff}$-[Fe/H] (b) and [Fe/H]-$\log{g}$ (c).

The micro-turbulent velocity is important when derived the lithium abundance, especially for the strong lines, thus, empirical relations have been adopted to determine this parameter. For giants ($\log{g}$ $\leq$ 3.5\,dex) with [Fe/H] higher than $-$1.0\,dex, the relation from \citet{Holtzman2018} is adopted, which was derived based on a metal-rich giant subsample of the APOGEE Data Releases 13:

\begin{equation}\label{vmic}
\emph{$\xi$} = 10^{(0.226-0.0228\log{g}+0.0297(\log{g})^2 -0.0113(\log{g})^3)}
\end{equation}

\noindent here, the unit of $\xi$ is in km\,s$^{-1}$. While, a linear relationship from \citet{Garca2016} is used for metal-poor ([Fe/H] $\leq -$1.0\,dex) giants:

\begin{equation}\label{vmic}
\emph{$\xi$} = 2.478 - 0.325 \log{g}
\end{equation}

For hot dwarfs ($T_{\rm eff}\textgreater$ 5000\,K, $\log{g}$ $\textgreater$ 3.5\,dex), we follow \citet{GaoX2018}'s suggestion, which has also been adopted for the GALAH survey \citep{Buder2020}:

\begin{equation}\label{vmic}
\emph{$\xi$} = 1.1 + 10^{-4} (T_{\rm eff} - 5500\,K) + 4 \times 10^{-7}(T_{\rm eff} - 5500\,K)^2
\end{equation}

While, a constant micro-turbulence of 1.0\,km\,s$^{-1}$ is set for dwarfs with $T_{\rm eff}$ lower than 5000\,K.

\section{METHOD}\label{sec:METHOD} 

The approach employed in this study is a revised method we used earlier to search for lithium-rich giants from the LAMOST low-resolution spectra \citep{Gao2019}. The method is based on template-matching, and we briefly describe the processes as follows.

\subsection{The synthetic spectra} 

The synthetic spectra have been calculated using the SPECTRUM synthesis code \citep[v2.77, 2017,][] {Gray1999} based on the ATLAS9 stellar atmosphere model from \citet{Castelli2003} under a LTE assumption, and the atomic line data for lithium were from \citet{shi2007}. The resolution was fixed as 0.5\,{\AA} for all of the synthetic spectra with the micro-turbulent velocities calculated by the empirical relations presented in Sect. \ref{sec:DATA}.

The grids of stellar parameters and lithium abundance of synthetic spectra were set as:

3,500\,K $\leq$ $T_{\rm eff}$ $\leq$ 4,500\,K in a step of 200\,K 

4,500\,K \textless $T_{\rm eff}$ $\leq$ 6,500\,K in a step of 100\,K

6,500\,K \textless $T_{\rm eff}$ $\leq$ 7,500\,K in a step of 200\,K

 0.0 $\leq$ $\log{g}$ $\leq$ 5.0 in a step of 0.5\, dex

 $-$2.6 \textless [Fe/H] $\leq$ 0.6 in a step of 0.2\,dex 

 $-$3.0 $\leq$ [Li/Fe] $\leq$ 5.1 in a step of 0.1\,dex

 For the synthetic spectra with [Fe/H] \textless $-$0.6, we enhanced the abundances of $\alpha$-elements by 0.4\,dex.
 
The synthetic spectra around the \ion{Li}{1} resonance line at 6708 {\AA} with four sets of stellar parameters are shown in Figure \ref{fig:2}. 

\subsection{The lithium abundances} 

The lithium abundances were estimated with a template-matching method, and the following steps have been used: 

1. {\bf Pre-processing}: 
two initial procedures were needed to be done. First, we corrected the wavelength of an observed spectrum to the rest frame scale using the radial velocity derived by cross-correlation. Second, we degraded the resolution of the synthetic spectra according to the observed ones.

\begin{figure*}[!t]
\includegraphics[scale=0.413]{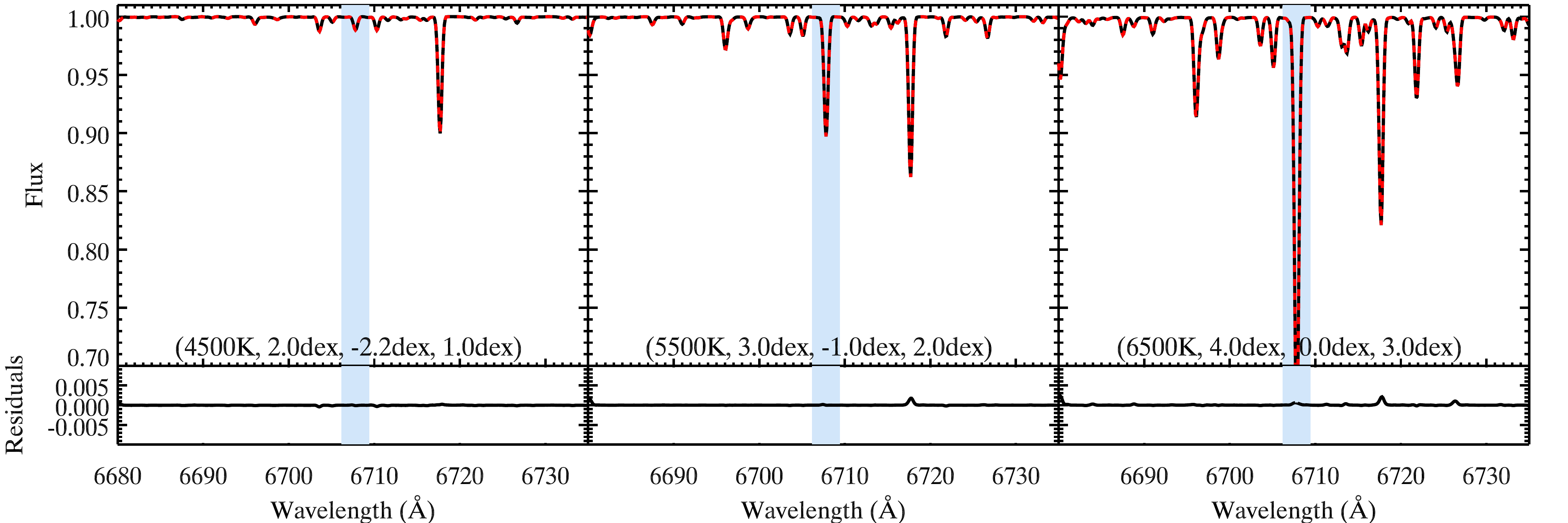}
\centering
\caption{Three cases of comparison between the synthetic spectra (black solid lines) and the interpolated ones (red dashed lines). The light blue region indicates the position of the \ion{Li}{1} resonance line. The stellar parameters are shown in the sequence of $T_{\rm eff}$, $\log{g}$, [Fe/H] and [Li/Fe], and the residuals are presented at the bottom of each panel.\label{fig:3}}
\end{figure*}

\begin{figure*}[!t]
\includegraphics[scale=0.435]{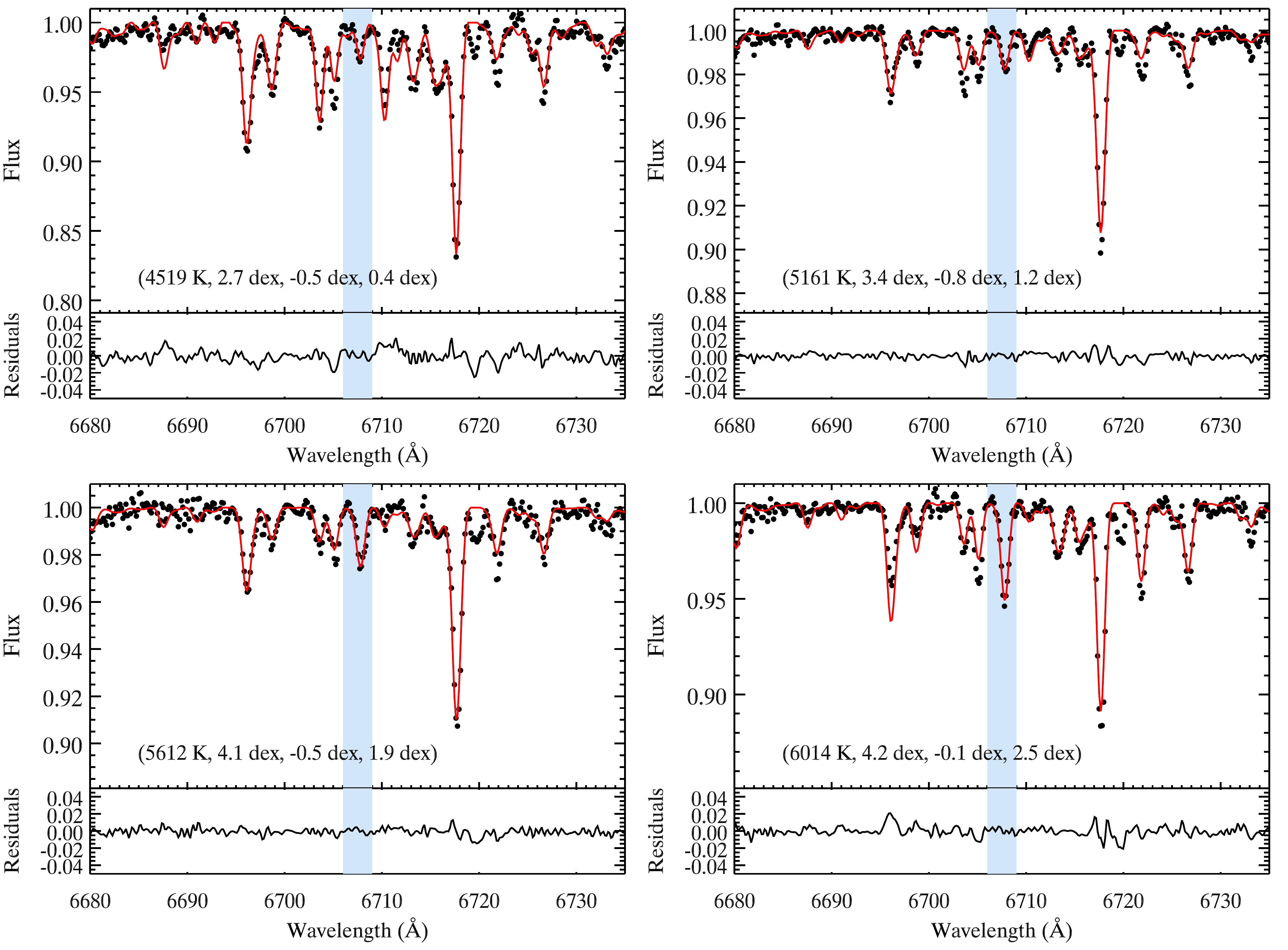}
\centering
\caption{Four matching examples. The black dotted lines and the red solid lines are the observed and the best-matching synthetic spectra, respectively. The wavelength range of 6706 to 6709\,{\AA} to compute $\chi^2$ is marked as light blue region. The stellar parameters ($T_{\rm eff}$, $\log{g}$, [Fe/H]) and A(Li) are presented, and the residuals are plotted at the bottom of each panel.\label{fig:4}}
\end{figure*}

2. {\bf Generating}: We generated a set of synthetic spectra by interpolating the adjacent grids of the synthetic spectra for the stellar atmospheric parameters of a target star, [Li/Fe] of the generating spectra were varying from $-$3.0 to $+$5.1. In order to examine the impact of the uncertainty introduced by our interpolation algorithm on the determination of lithium abundances, we compared the synthetic spectra with the interpolated ones for the same stellar parameters and lithium abundances for three cases as shown in Figure \ref{fig:3}. It was found that the differences were less than $1\%$, which had little effects on our results.

3. {\bf Calculating}: We computed chi-square ($\chi^2$) between the observed and each synthetic spectra around the wavelength range of 6706\,-\,6709\,{\AA}, which covered the \ion{Li}{1} resonance line. The $\chi^2$ is defined as

\begin{equation}
{\chi}^2=\sum_{i=1}^{N} \frac{(O_i - S_i)^2}{{\sigma}_i^2}
\end{equation}

where ${O_i}$ and ${S_i}$ are the flux of the ${i_{th}}$ point of observed and synthetic spectra respectively, ${\sigma}_i$ is the error of the ${i_{th}}$ observed flux, and N is the number of points used in calculation. Before calculating $\chi^2$, it is needed to adjust the continuums of the synthetic and observed spectra to a same level. We calculate the flux ratio between the observed and synthetic spectra in the wavelength range from 6675 to 6740\,{\AA} (the \ion{Li}{1} resonance line has been masked), and fit this set of ratio with a second-order polynomial. Finally, the continuum of a synthetic spectrum is adjusted to the level of the observed one by multiplying the fit values.

4. {\bf Estimating}: We fitted the $\chi^2$ array with a third-order polynomial to determine the minimum $\chi^2$, and the [Li/Fe] can be determined. Then we transformed [Li/Fe] into the commonly used form A(Li)\footnote{A(Li) $= \log({\rm N_{Li}}/{\rm N_{H}}) + 12$, where ${\rm N_{Li}}$ and ${\rm N_{H}}$ is the number density of lithium and hydrogen, respectively.}: A(Li) = [Li/Fe] + [Fe/H] + A(Li){$_\odot$}, here A(Li){$_\odot$} was adopted as 1.15\,dex according to the value used in SPECTRUM code.

Figure \ref{fig:4} shows four matching examples, the black dotted lines and red solid lines are the observed and the best-matched synthetic spectra, respectively. The residuals plotted at the bottom of each panel indicate that the deviations in the considered region between the observed and the best-matched spectrum are less than one percent. 

\begin{figure*}
\includegraphics[scale=0.6]{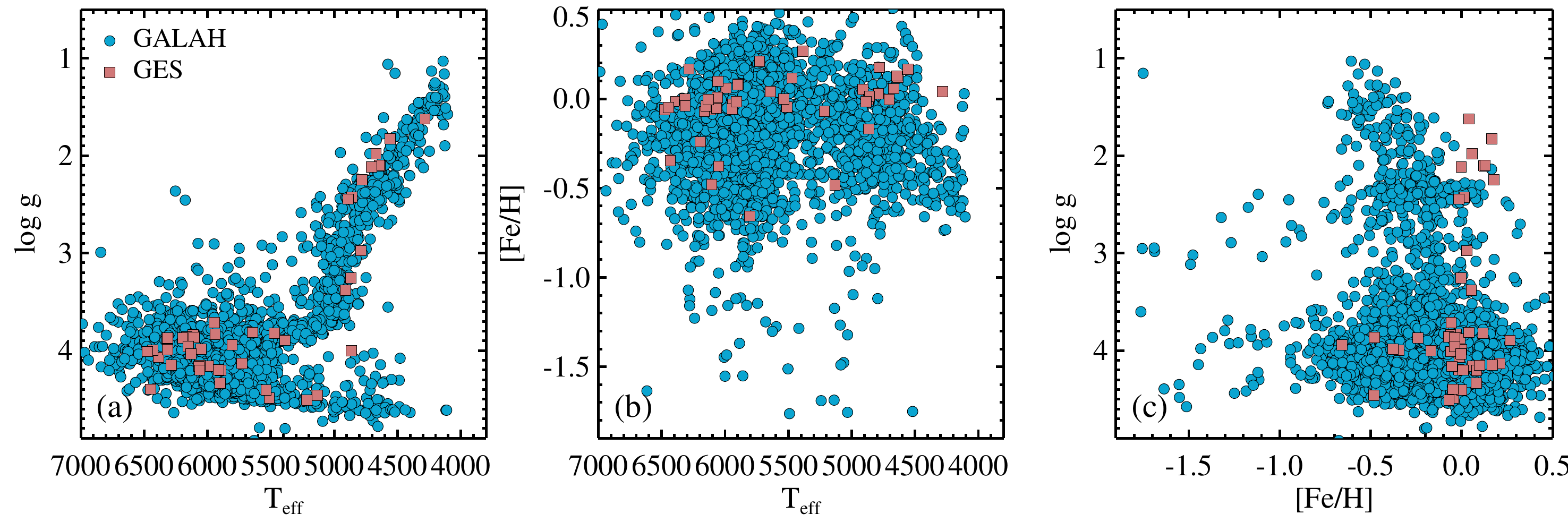}
\centering
\caption{Distribution of the atmospheric parameters for the common sources. The Blue dots and orange squares indicate the GALAH and GES targets, respectively.
\label{fig:5}}
\end{figure*}

\begin{figure*}
\includegraphics[scale=0.43]{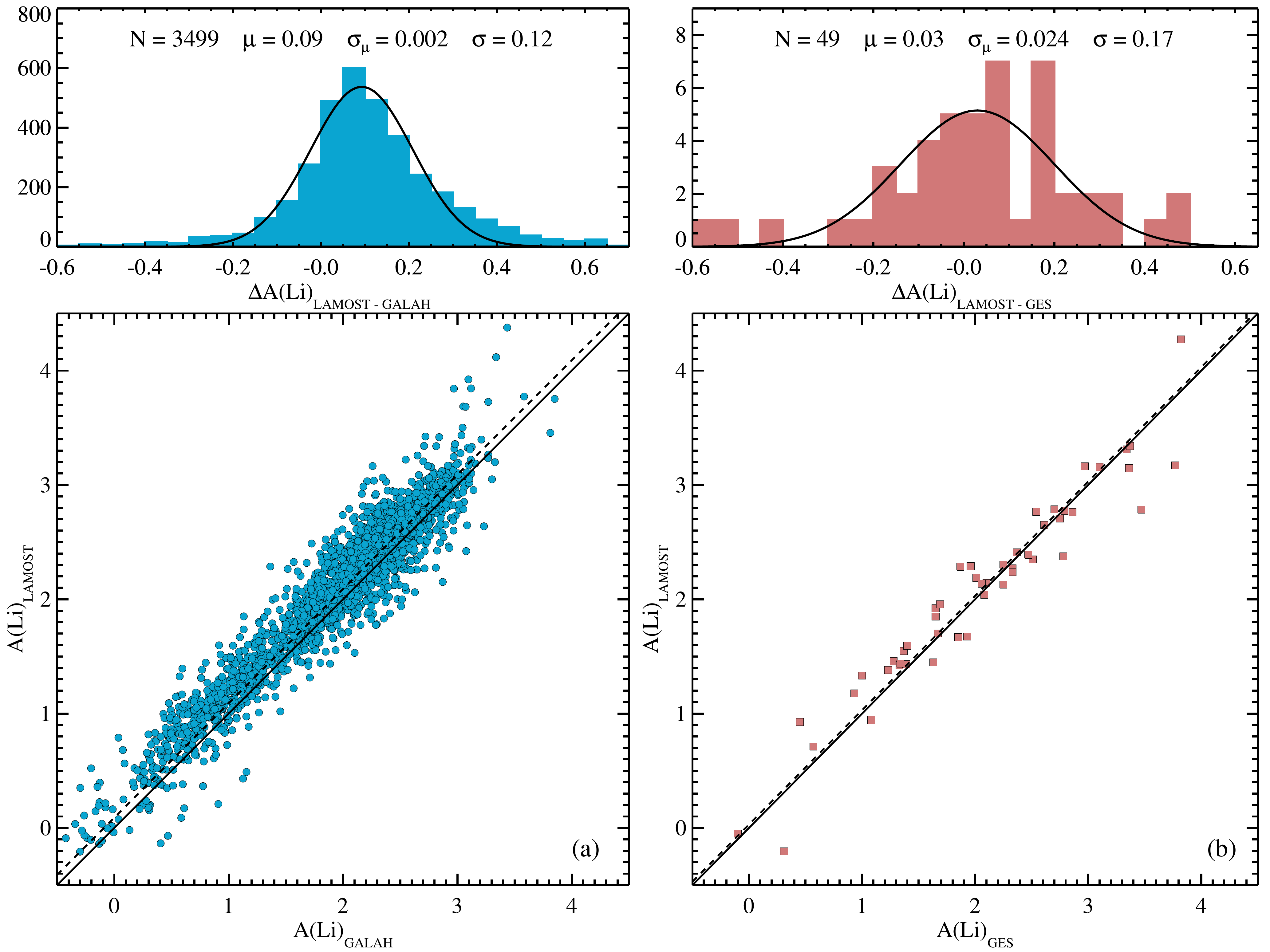}
\centering
\caption{Comparisons of the lithium abundances calculated by our method (A(Li)$_{\rm LAMOST}$) and these determined from GALAH (A(Li)$_{\rm GALAH}$) (a) and GES (A(Li)$_{\rm GES}$) (b). The black solid lines are the one-to-one correspondence, while the black dashed lines are the overall shifts. The distribution of the differences between A(Li)$_{\rm LAMOST}$ and A(Li)$_{\rm GALAH, GES}$ with the Gaussian fitting result is plotted on the top of each panel, and the number of targets (N), the mean value ($\mu$), the uncertainty of the mean ($\sigma_\mu$) and the dispersion ($\sigma$) are presented. The meanings of symbols are the same as in Figure\,\ref{fig:5}.
\label{fig:6}}
\end{figure*}

\begin{figure*}
\includegraphics[scale=0.7]{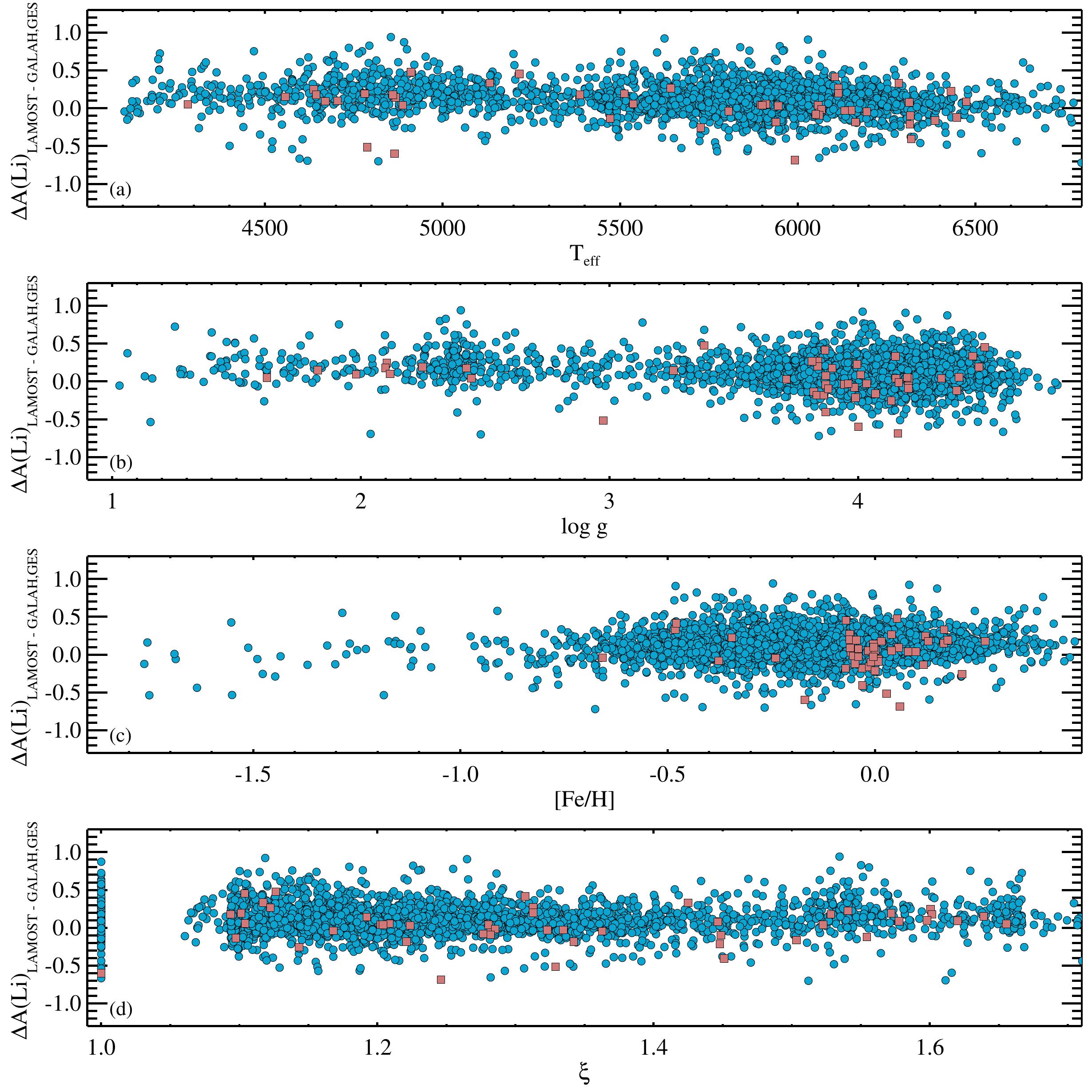}
\centering
\caption{The differences of A(Li) determined by us with those from the high resolution spectra (A(Li)$_{\rm LAMOST}- {\rm A(Li)}_{\rm GALAH,GES})$ as functions of $T_{\rm eff}$ (a), $\log{g}$ (b), [Fe/H] (c) $\xi$ (d). The micro-turbulences are calculated by the empirical relations presented in Sect. \ref{sec:DATA}. The meanings of symbols are the same as in Figure\,\ref{fig:5}.
\label{fig:7}}
\end{figure*}

\section{Validation of the method} \label{sec:Validation}

It is necessary to validate our method used for calculating the lithium abundances, therefore, we compared A(Li) derived by us with those from the high resolution spectral analysis for the common stars in GALAH DR3 \citep[GALAH Data Release 3,][GALAH hereafter]{Buder2020} and Gaia-ESO Survey iDR4 (the fourth internal data release of the Gaia-ESO Survey, the catalogue is available on the ESO Phase 3 webpage\footnote{http://www.eso.org/qi/}, GES hereafter). There are 3,499 targets in GALAH (for flag$_{\rm [Li/Fe]}$=0) and 49 in GES have also been observed by the LAMOST MRS, respectively. The distribution of stellar parameters of these targets is shown in Figure\,\ref{fig:5}. 

To avoid the difference caused by adopting different stellar parameters in each sample, we adopted $T_{\rm eff}$, $\log{g}$ and [Fe/H] from GALAH and GES in our calculation, and the comparisons of A(Li) estimated by our method with these from GALAH and GES are presented in Figure\,\ref{fig:6}. We found a good agreement with an offset of +0.09\,dex and a scatter of 0.12\,dex for the GALAH common stars (panel a). The uncertainty of this offset is 0.002\,dex, which indicates that +0.09\,dex is a statistically significant offset. It need to be pointed out that the lithium abundances provided by GALAH are based on non-LTE, while GES and ours are based on a LTE assumption. \citet{Amarsi2020} presented that the non-LTE effects for the \ion{Li}{1} resonance lines are negative with a typical value of $\Delta$A(Li)$_{\rm (non-LTE-LTE)} =-0.10$\,dex for dwarf stars. The majority of the common GALAH samples are dwarfs, therefore, the offset can be explained by the non-LTE effects. The comparison with GES shows an offset of +0.03\,dex with a scatter of 0.17\,dex (panel b).  The uncertainty of this offset is 0.024\,dex. We also investigate the differences of A(Li) as functions of each stellar parameters and find that there is no obvious correlation for any parameter as shown in Figure\,\ref{fig:7}.

\section{Error estimation}\label{sec:Error}

 \begin{figure*}
 \includegraphics[scale=0.41]{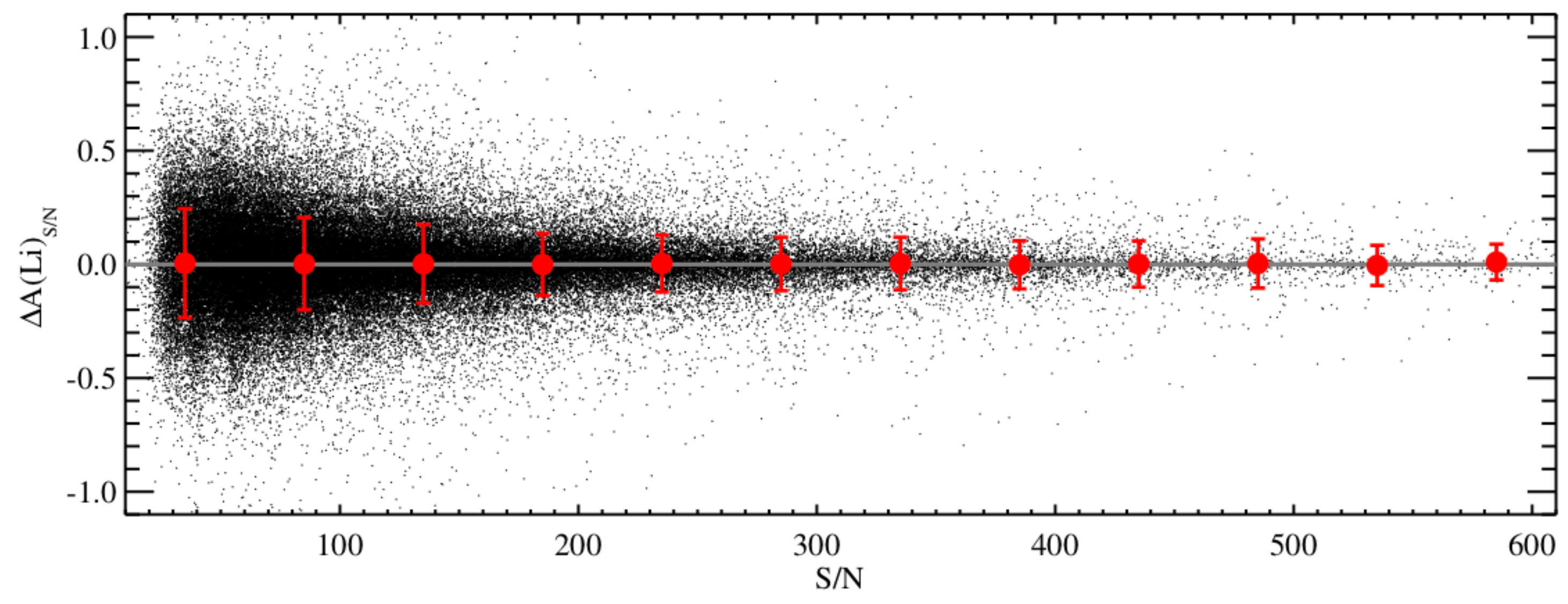}
 \centering
 \caption{The differences of A(Li) between two epoch observations as a function of S/N. The red dots and the vertical error bars represent the mean and the standard deviation, respectively. The bin size is 50.
 \label{fig:8}}
 \end{figure*}

 \begin{figure*}
 \includegraphics[scale=0.7]{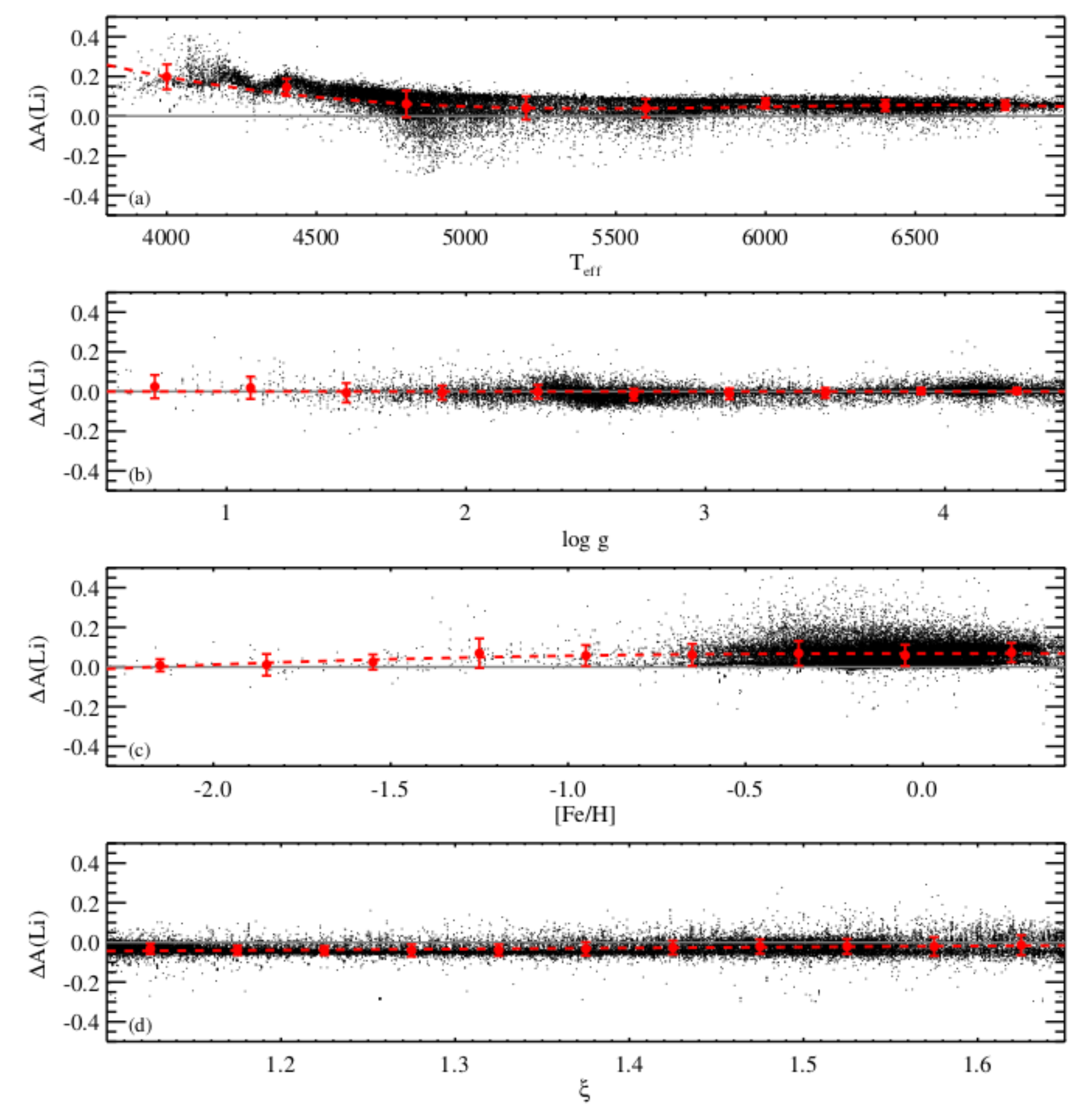}
 \centering
 \caption{ The deviations of the lithium abundances derived by changing stellar parameters with $\Delta T_{\rm eff} = +100\,{\rm K}$ (a), $\Delta \log{g} = +0.25$\,dex (b), $\Delta$[Fe/H] = $+0.1$\,dex (c) and $\Delta$$\xi$ = $+0.5$ km\,s$^{-1}$ (d) as functions of their corresponding parameters. Red dots and vertical error bars are the mean and the standard deviation, and the bin sizes are 400\,K, 0.4\,dex, 0.3\,dex and 0.05 km\,s$^{-1}$, respectively. The red dashed lines are the fit to the mean values with third-order polynomial (Panels a, b and c) and first-order polynomial (Panel d). The micro-turbulences are calculated with the empirical relations presented in Sect. \ref{sec:DATA}. Only results from spectra with S/N$>$ 200 are adopted.
 \label{fig:9}}
 \end{figure*}

\begin{figure}[!t]
\centering
\includegraphics[scale=0.28]{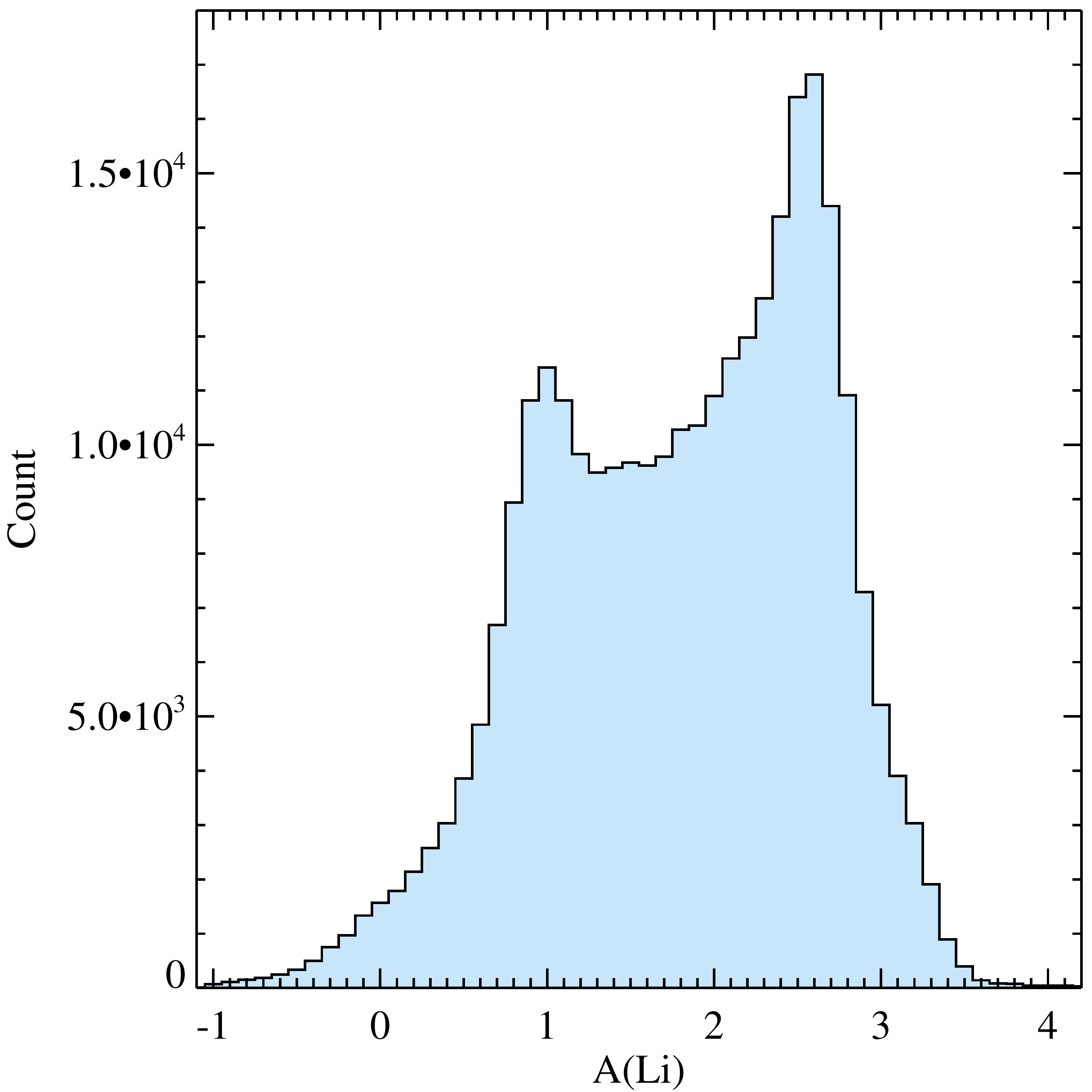}
\caption{Histogram of A(Li) for our sample. The bin size is 0.1\,dex. \label{fig:10}}
\end{figure}

\begin{figure*}
\includegraphics[scale=0.59]{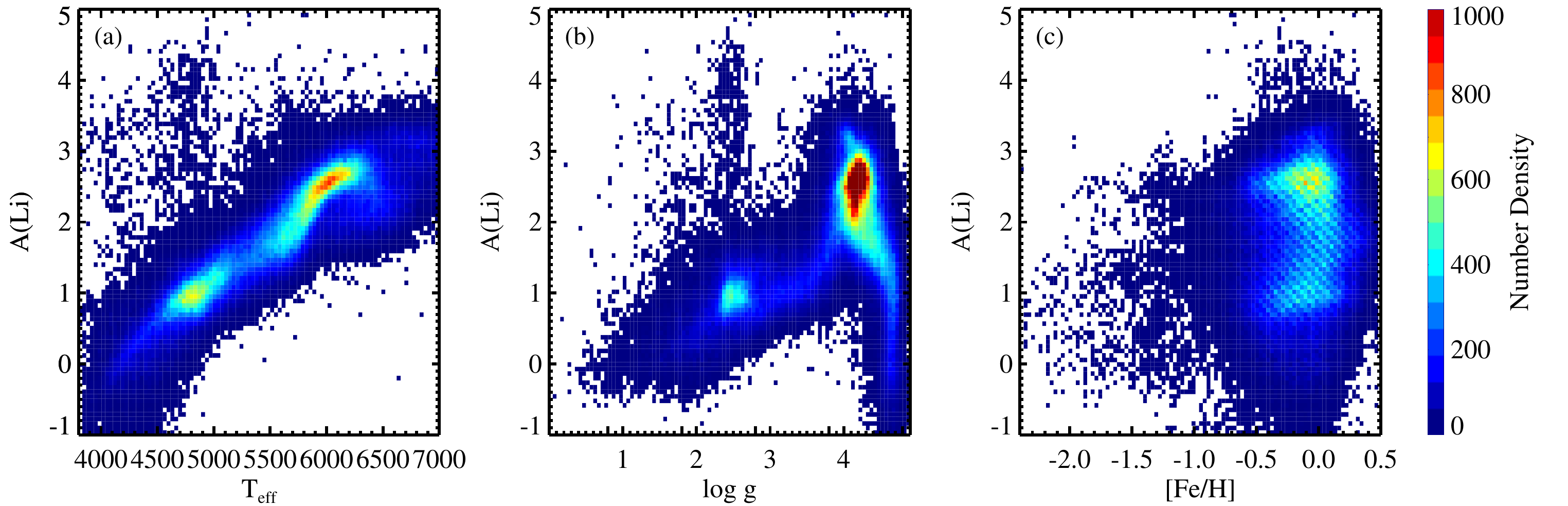}
\centering
\caption{The lithium abundances versus $T_{\rm eff}$ (a), $\log{g}$ (b) and [Fe/H] (c). \label{fig:11}}
\end{figure*}

The quality of the observed spectra and the uncertainties in stellar parameters will influence the accuracy of lithium abundance measurement, and we discuss the errors caused by these factors as follows.

\subsection{Errors due to the quality of spectra}

Taking advantage of the repeated observations for the same stars, we estimate the errors of lithium abundances related to the quality of spectra. To do this, $43,033$ targets having repeated observation with the differences in spectral S/N less than $20\%$ have been selected. Figure \ref{fig:8} shows the difference in A(Li) between two observed spectra of a target as a function of S/N. The results indicate that the differences in A(Li) measured from the spectra are sensitive to their S/Ns, which is reasonable. We note that the dispersions decrease from 0.3 to 0.1\,dex with increasing S/N, and they can be fitted with a first-order polynomial.

\begin{equation}\label{sn}
\Delta{\rm A(Li)_{\rm S/N}} = a_{0} + a_{1} \times \rm S/N
\end{equation}

The fitting coefficients, $a_{0}$ and $a_{1}$, can be found in Table \ref{chartable1}.

\begin{deluxetable}{llll}
\tablecaption{Coefficients of the polynomial \ref{sn}, \ref{teff} \ref{feh} and \ref{evmic}.
\label{chartable1}}
\startdata 
 & &   \\                 
a$_{0}$ & a$_{1}$   \\             
2.09(-1) & -2.44(-4)  \\ 
\hline           
b$_{0}$ & b$_{1}$ & b$_{2}$ & b$_{3}$ \\             
5.14 & -2.58(-3) & 4.33(-7) & -2.39(-11)\\ 
\hline            
c$_{0}$ & c$_{1}$ & c$_{2}$ & c$_{3}$ \\             
6.80(-2) & -8.54(-4) & -1.01(-2) & 2.18(-3) \\             
\hline            
d$_{0}$ & d$_{1}$\\             
-1.00(-1) & 5.13(-2)\\             
\enddata                  
\tablecomments{The integers in parentheses refer to powers of 10, for example, 2.09(-1) represents 2.09 $\times$ 10$^{-1}$.}
\end{deluxetable} 

\subsection{Errors due to the uncertainties in the stellar parameters}

The uncertainties in atmospheric parameters will introduce errors in our measurements, and the following analysis was performed separately for the two sets of stellar parameters adopted.
 
\subsubsection{The LASP-SP}

The LASP-SP have typical uncertainties of 100\,K, 0.25\,dex and 0.10\,dex for the effective temperature, surface gravity and metallicity, respectively \citep{Luo2015}, and the typical uncertainty in micro-turbulent velocity calculated with the empirical relations is around 0.5 km\,s$^{-1}$. To evaluate the influence of the uncertainties of stellar parameters on the lithium abundances, we recalculated them with a change of the effective temperature by $+$100\,K, the surface gravity by $+$0.25\,dex, the metallicity by $+$0.10\,dex and the micro-turbulence velocity by $+0.5$ km\,s$^{-1}$ for the spectra with S/N $\textgreater$ 200, respectively. Variations of the obtained lithium abundances for the change of stellar parameters are shown in Figure \ref{fig:9}. 

It is clear that the errors are large for cool stars, which may be caused by the stronger \ion{Li}{1} resonance line at lower $T_{\rm eff}$ (Panel a). The uncertainties in $\log{g}$ will introduce negligible errors in the lithium abundance calculation (Panel b). While, the errors for metal-poor stars are smaller than those for metal-rich targets (Panel c), which are due to the influence of a nearby \ion{Fe}{1} line. The uncertainties in $\xi$ result in small errors in the derived lithium abundances (Panel d). We fitted the mean values in these bins with a third-order polynomial as functions of $T_{\rm eff}$ and [Fe/H], while a first-order polynomial for $\xi$, respectively.

\begin{equation}\label{teff}
\Delta{\rm A(Li)_{T_{\rm eff}}}= b_{0} + b_{1} \times T_{\rm eff} + b_{2} \times T_{\rm eff}^2 + b_{3} \times T_{\rm eff}^3
\end{equation}

\begin{equation}\label{feh}
\begin{aligned}
\Delta{\rm A(Li)_{\rm [Fe/H]}} = &~c_{0} + c_{1} \times {\rm [Fe/H]} + c_{2} \times {\rm [Fe/H]}^2 \\
                                                 &+ c_{3} \times {\rm [Fe/H]}^3
\end{aligned}
\end{equation}

\begin{equation}\label{evmic}
\begin{aligned}
\Delta{\rm A(Li)_{\xi}} = d_{0} + d_{1} \times \xi
\end{aligned}
\end{equation}

The fitting coefficients are also listed in Table \ref{chartable1}.

Therefore, the total error of lithium abundance for every spectrum in LASP-SP can be estimated as

\begin{equation}\label{lasp}
\begin{aligned}
\Delta{\rm A(Li)} =\\ \sqrt{\Delta{\rm A(Li)_{\rm S/N}}^2 + \Delta{\rm A(Li)_{T_{\rm eff}}}^2 + \Delta{\rm A(Li)_{\rm [Fe/H]}}^2 + \Delta{\rm A(Li)_{\xi}}^2}
\end{aligned}
\end{equation}

\subsubsection{The APOGEE-SP}

The APOGEE-SP are provided by APOGEE DR16, the mean errors of $T_{\rm eff}$, $\log{g}$ and [Fe/H] are 100\,K, 0.07\,dex and 0.01\,dex, respectively \citep{Jonsson2020}. Thus the errors of lithium abundances due to the uncertainties in $T_{\rm eff}$ and $\xi$ are similar to these of the LASP-SP, while the influences of uncertainties in $\log{g}$ and [Fe/H] are negligible. The total error of lithium abundance for every spectrum in APOGEE-SP is estimated as 

\begin{equation}
\Delta{\rm A(Li)} = \sqrt{\Delta{\rm A(Li)_{\rm S/N}}^2 + \Delta{\rm A(Li)_{T_{\rm eff}}}^2 + \Delta{\rm A(Li)_{\xi}}^2}
\end{equation}

\section{Results and discussions}\label{sec:RESULT}

The lithium abundances derived from LAMOST MRS spectra by us are presented in Table \ref{chartable2}. Columns (1) and (2) of Table \ref{chartable2} are the LAMOST\_ids and obs\_ids, respectively. The {\it Gaia} DR2 source\_ids are given in Column (3). Columns (6) - (8) are the stellar parameters. The S/Ns of the spectra are shown in column (9), while the lithium abundances and their errors are given in columns (10) and (11), the typical uncertainty is $\sim$ 0.2\,dex. The final column contains the sources of stellar parameters . 

We present the histogram of lithium abundances in Figure \ref{fig:10}, while the lithium abundances versus the stellar parameters are plotted in Figure \ref{fig:11}. The distribution of the lithium abundance vs. $T_{\rm eff}$ indicates that the detection limit of lithium abundance is sensitive to $T_{\rm eff}$, and this limit is about $-$1.0\,dex for the coolest stars in our sample. Also, the lithium abundances are strongly correlated to the temperature for the majority of stars as shown in Figure \ref{fig:11}(a), a similar phenomenon can be found from the GALAH DR3 \citep{Buder2020}. The histogram presents two clear peaks around $+$2.6\,dex and $+$1.0\,dex, respectively. The former one at $+$2.6\,dex is dominated by hot dwarfs, which indicates the minimum of the stellar Li depletion for F- and G-dwarfs, this phenomenon has also been found in open clusters \citep{Thorburn1993, Anthony-Twarog2009, Cummings2017} and field stars. The latter one at $+$1.0\,dex is dominated by giants, which is consistent with the discovery reported in open clusters \citep{Gilroy1989, Anthony-Twarog2021}.  This result support that a more severe lithium depletion happens in giant phase. While some dwarfs have lithium abundance even higher than that of the meteorites, \citet{2002Deliyannis} suggested that the upward diffusion may enrich the surface lithium in some dwarfs under a narrow effective temperature range. There are a small fraction of giants with lithium abundances higher than 1.5\,dex, which are denominated as lithium-rich giants, most of them have $\log{g} \sim 2.4\,dex$. This result suggests the existence of lithium enrichment mechanisms in some giants \citep[][]{Casey2016}.

\section{SUMMARY}\label{sec:SUMMARY}
In this work, we provide the lithium abundances for 294,857 spectra of 165,479 stars. We derived the lithium abundances from LAMOST MRS spectra using a template-matching method. By comparing the lithium abundances determined by our method with those from the high resolution surveys including \emph{Gaia}-ESO and GALAH, a good consistence can be found. The typical error of our lithium abundances is $\sim$ 0.2\,dex. The lithium abundance derived in this work will be helpful for understanding the physical mechanisms occurring inside stars that  either deplete the surface stellar lithium or create and enhance the surface stellar lithium, which will be investigated in our forthcoming works.

\begin{longrotatetable}
\begin{deluxetable*}{ccrcccccrccc}
\tablecaption{The information and resulting lithium abundances of our sample. (This table is available in its entirety in machine-readable form.)\label{chartable2}}
\tablewidth{700pt}
\tablehead{
\colhead{LAMOST\_id} & \colhead{ obs\_id} & \colhead{ {\it Gaia} DR2 source\_id} & \colhead{R.A.} & \colhead{Decl.} & \colhead{$T_{\rm eff}$} & 
\colhead{$\log{g}$} & \colhead{[Fe/H]} & \colhead{S/N} & \colhead{A(Li)} & \colhead{$\delta{\rm A(Li)}$} & \colhead{para\_source$^{\rm a}$} \\ 
\colhead{} & \colhead{} & \colhead{} & \colhead{h:m:s (J2000)} & \colhead{d:m:s (J2000)} & \colhead{(K)} & \colhead{(dex)} & \colhead{(dex)} & \colhead{} & 
\colhead{({\it dex})} &\colhead{({\it dex})}} 
\startdata
  LAMOST J000000.32+573710.2 &    695107025 &    422596679964513792  &    00:00:00.32 &    +57:37:10.2 &     6407  &     4.0   &     -0.1  &     121   &     2.4   &     0.2   &     L     \\
  LAMOST J000000.96+411611.6 &    596808097 &    2882275590828656384 &    00:00:00.96 &    +41:16:11.6 &     5155  &     3.6   &     -0.3  &     241   &     1.0   &     0.2   &     L     \\
  LAMOST J000011.90+003841.9 &    612410091 &    2738294680609708032 &    00:00:11.90 &    +00:38:41.9 &     4624  &     4.5   &     -0.1  &     62    &     0.3   &     0.2   &     L     \\
  LAMOST J000026.11+042646.9 &    682016169 &    2740418627837183360 &    00:00:26.11 &    +04:26:46.9 &     5462  &     4.6   &     0.3   &     38    &     1.6   &     0.2   &     L     \\
  LAMOST J000032.76+002107.6 &    695916238 &    2738205104771771648 &    00:00:32.76 &    +00:21:07.6 &     5372  &     4.3   &     -0.8  &     49    &     1.3   &     0.2   &     L     \\
  LAMOST J000032.76+002107.6 &    695916238 &    2738205104771771648 &    00:00:32.76 &    +00:21:07.6 &     5372  &     4.3   &     -0.8  &     49    &     1.3   &     0.2   &     L     \\
  LAMOST J000034.04+095225.0 &    694503069 &    2765223953757033856 &    00:00:34.04 &    +09:52:25.0 &     5490  &     4.1   &     0.4   &     68    &     2.0   &     0.2   &     L     \\
  LAMOST J000035.84+090210.1 &    694510214 &    2747114791089408768 &    00:00:35.84 &    +09:02:10.1 &     6080  &     4.4   &     -0.2  &     136   &     2.6   &     0.2   &     L     \\
  LAMOST J000042.90+005734.1 &    612410125 &    2738310039412786944 &    00:00:42.90 &    +00:57:34.1 &     6084  &     4.1   &     -0.4  &     36    &     3.2   &     0.2   &     L     \\
  LAMOST J000042.90+005734.1 &    612410125 &    2738310039412786944 &    00:00:42.90 &    +00:57:34.1 &     6084  &     4.1   &     -0.4  &     36    &     3.2   &     0.2   &     L     \\
  LAMOST J000043.49+091720.6 &    694503241 &    2747173030845956480 &    00:00:43.49 &    +09:17:20.6 &     5685  &     4.3   &     -0.8  &     228   &     1.1   &     0.2   &     L     \\
  LAMOST J000044.10+044910.7 &    682016057 &    2741942516593385216 &    00:00:44.10 &    +04:49:10.7 &     5202  &     3.9   &     0.1   &     129   &     1.4   &     0.2   &     L     \\
  LAMOST J000044.44+544016.3 &    609207025 &    420369790961162496  &    00:00:44.44 &    +54:40:16.3 &     5799  &     4.4   &     -0.3  &     98    &     2.1   &     0.2   &     L     \\
  LAMOST J000044.78+090400.0 &    694510207 &    2747115306485484032 &    00:00:44.78 &    +09:04:00.0 &     6296  &     4.0   &     0.0   &     271   &     1.8   &     0.2   &     L     \\
  LAMOST J000045.04+085702.9 &    694510201 &    2747109671488394624 &    00:00:45.04 &    +08:57:02.9 &     5649  &     4.4   &     -0.3  &     79    &     1.4   &     0.2   &     L     \\
  LAMOST J000045.72+411411.5 &    598108078 &    2882277407599002368 &    00:00:45.72 &    +41:14:11.5 &     5826  &     4.2   &     0.1   &     41    &     2.0   &     0.2   &     L     \\
  LAMOST J000045.72+411411.5 &    598108078 &    2882277407599002368 &    00:00:45.72 &    +41:14:11.5 &     5826  &     4.2   &     0.1   &     41    &     2.0   &     0.2   &     L     \\
  LAMOST J000046.05+100534.5 &    694503109 &    2765263188281968384 &    00:00:46.05 &    +10:05:34.5 &     5809  &     4.1   &     0.2   &     120   &     2.4   &     0.2   &     L     \\
  LAMOST J000047.56+003056.2 &    612410177 &    2738242178929502592 &    00:00:47.56 &    +00:30:56.2 &     4973  &     4.4   &     0.3   &     89    &     1.0   &     0.2   &     A     \\
  LAMOST J000047.79+081951.0 &    694502035 &    2746863312164361088 &    00:00:47.79 &    +08:19:51.0 &     4820  &     2.9   &     -0.4  &     225   &     0.8   &     0.2   &     L     \\
  LAMOST J000051.36+410852.9 &    597408085 &    2882223462809766272 &    00:00:51.36 &    +41:08:52.9 &     6194  &     4.3   &     0.0   &     80    &     2.9   &     0.2   &     L     \\
  LAMOST J000052.60+573549.7 &    596909063 &    422595889690565248  &    00:00:52.60 &    +57:35:49.7 &     4182  &     1.7   &     -0.1  &     154   &     0.2   &     0.2   &     L     \\
  LAMOST J000052.60+573549.7 &    596909063 &    422595889690565248  &    00:00:52.60 &    +57:35:49.7 &     4182  &     1.7   &     -0.1  &     154   &     0.2   &     0.2   &     L     \\
  LAMOST J000053.43+004059.4 &    612410090 &    2738248909142217600 &    00:00:53.43 &    +00:40:59.4 &     5840  &     4.2   &     -0.3  &     177   &     2.4   &     0.2   &     A     \\
... & ... & ... & ... & ... & ... & ... & ... &... & ... & ... & ...\\\enddata
\tablecomments{$^{\rm a}$the sources of stellar parameters, `L' represents LASP and `A' represents APOGEE.}
\end{deluxetable*}
\end{longrotatetable}

\emph{Acknowledgements.} 
The authors thank the anonymous referee for the comments,  which have helped to improve the manuscript. Our research is supported by National Key R\&D Program of China No.2019YFA0405502, the National Natural Science Foundation of China under grant Nos. 12090040, 12090044, 11833006, 12022304, 11973052, 11973042 and U1931102. This work is supported by the Astronomical Big Data Joint Research Center, co-founded by the National Astronomical Observatories, Chinese Academy of Sciences and Alibaba Cloud. H.-L.Y. acknowledges support from the Youth Innovation Promotion Association of the CAS (id. 2019060). This work is also partially supported by the Open Project Program of the Key Laboratory of Optical Astronomy, National Astronomical Observatories, Chinese Academy of Sciences. Guoshoujing Telescope (the Large Sky Area Multi-Object Fiber Spectroscopic Telescope LAMOST) is a National Major Scientific Project built by the Chinese Academy of Sciences. Funding for the project has been provided by the National Development and Reform Commission. LAMOST is operated and managed by the National Astronomical Observatories, Chinese Academy of Sciences. We also used the data from APOGEE (https://www.sdss.org/dr16/irspec/), Gaia-ESO(http://www.eso.org/qi/) and GALAH (https://galah- survey.org/).

\emph{Software:} TOPCAT \citep{Taylor2005}, SPECTRUM \citep{Gray1999}.

\end{document}